# Modeling Web Evolution

Michalis Vafopoulos[*], Efstathios Amarantidis, Ioannis Antoniou

Aristotle University of Thessaloniki, Mathematics Department, Graduate Program in Web Science

**Abstract**

The Web is the largest human information construct in history transforming our society. How can we understand, measure and model the Web evolution in order to design effective policies and optimize its social benefit? Early measurements of the Internet traffic and the Web graph indicated the scale-free structure of the Web and other Complex Networks. Going a step further Kouroupas, Koutsoupias, Papadimitriou and Sideri (KKPS) presented an economic-inspired model which explains the scale-free behavior as the interaction of Documents, Users and Search engines. The purpose of this paper is to clarify the open issues arising within the KKPS model through analysis and simulations and to highlight future research developments in Web modeling, which is the backbone of Web Science.

## 1. Introduction

In the first twenty years of its existence, the World Wide Web (Web or www) has proven, to have had a fundamental and transformative impact on all facets of our society. While the Internet has been introduced 20 years earlier, the Web has been its "killer" application with more than 1.5 billion users worldwide accessing more than 1 trillion web pages (excluding those that cannot be indexed, the "Deep Web") [1]. Searching, social networking, video broadcasting, photo sharing, blogging and micro-blogging have become part of everyday life whilst the majority of software and business applications have migrated to the Web. Today, the enormous impact, scale and dynamism of the Web in time and space demand more than our abilities to observe and measure its evolution process. Quantifying and understanding the Web lead to Web modeling, the backbone of Web Science research [2]. Web models should invest on Complexity [3] beyond reductionism, linking structure to function and evolution [4, 5, 6]. In this context, causality between events, temporal ordering of interactions and spatial distribution of Web components are becoming essential to addressing scientific questions at the Web techno-social system level.

---

[*] Corresponding author. E-mail address: vaf@webscience.gr, Aristotle University of Thessaloniki, Mathematics Department, 541 24, Thessaloniki, Greece. Tel.: +306945897894, fax: +302 310997929.

[1]

The first steps towards understanding cyberspace involved measurements and statistics of the Internet traffic [7, 8, 9, 10, 11]. The self-similar feature of the Internet was also found in the Web through preferential attachment [12, 13, 14]. Network science being a useful mathematical framework to formulate the non-reducible interdependence of Complex Systems [4, 5] recently led to significant results not only in Web graph statistics, but moreover in biology [15], economics [16] and sociology [17]. These results initiated a new understanding of Complexity in nature [18].

The statistical analysis of the Web graph led to four major findings [19]: *on-line property* (the number of nodes and edges changes with time), *power law degree distribution with an exponent bigger than 2*, *small world property* (the diameter is much smaller than the order of the graph) and *many dense bipartite subgraphs*.

In the light of these findings Kouroupas, Koutsoupias, Papadimitriou and Sideri proposed an economic-inspired model of the Web (KKPS model thereafter) [20, 21] which explains the scale-free behavior. Web evolution is modeled as the interaction of Documents, Users and Search Engines. The Users obtain satisfaction (Utility), when presented with some Documents by a Search Engine. The Users choose and endorse Documents with highest Utility and then the Search Engines improve their recommendations taking into account these endorsements, but not the dynamic interdependence of the Utility on the www state. Commenting on their results the authors have pointed out that (A) "more work is needed in order to define how the various parameters affect the exponent of the distribution" (of the in-degree of documents) and that (B) "increasing *b (the number of endorsed documents)* causes the efficiency of the algorithm to decrease. This is quite unexpected, since more user endorsements mean more complete information and more effective operation of the search engine. But the opposite happens: more endorsements per user seem to confuse the search engine."

The purpose of this paper is to address and clarify the issues (A), (B) arising within the Kouroupas, Koutsoupias, Papadimitriou and Sideri [21] modeling scheme (KKPS), through analysis and simulations and to highlight future research developments in Web modeling. In Section 2 we present the KKPS model. The results of our simulations on the power-law exponent and on the efficiency of the Search Algorithm are described in Section 3 and 4. After discussing the necessary modifications and extensions towards more realistic models for Web evolution in Section 5, we present our conclusions in Section 6.

## 2. The KKPS Web model

### 2.1. Assumptions

In the Web three types of entities are distinguished, namely: the Documents (i.e. web pages, created by authors), the Users and the Topics. The corresponding numbers are denoted by $n$, $m$ and $k$. It is assumed that:

$$k \leq m \leq n \tag{1}$$

The relation among Documents, Users and Topics is explained by the following statements [21]: "A User obtains satisfaction (Utility) if he or she is presented with some Documents by a Search Engine. The Search Engine recommends Documents to the Users, Users choose and endorse those that have the highest Utility for them, and then Search Engines make better recommendations based on these endorsements". "For each topic $t \leq k$ there is a Document vector $D_t$ of length $n$, with entries drawn independently from some distribution. The value 0 is very probable so that about $k - 1$ of every $k$ entries are 0".

In order to be more specific we shall denote by $D_{td}$ the Document components which represent the relevance of Document $d$ for Topic $t$ where $t = 1,..., k$ is the Topic index and $d=1,..., n$ is the Document index. "There are Users that can be thought as simple *Queries* asked by individuals". "For each topic $t$ there is a User vector $R_t$ of length $m$, whose entries also follow the distribution $Q$, with about $m/k$ non-zero entries" [21]. We understand the above statement as an attempt to simplify the User-Query interplay. In order to be more specific we shall denote by $R_{ti}$ the components of $R_t$ which represents the relevance of User-Query $i$ for Topic $t$, where $t = 1,..., k$ is the Topic index and $i = 1,..., m$ is the User-Query index. The Documents are linked with the User-Query set through Search Engines. Each Search Engine indexes Documents according to an algorithm and provides Users-Queries with Document recommendations based on information it has about their preferences for specific Topics. The Search Engine initially has no knowledge of Utility, but acquires such knowledge only by observing User-Query endorsements. In the ideal situation in which the Search Engine knows Utility, it would work with perfect efficiency, recommending to each User-Query the Documents he or she likes the most [21]. The Utility associated with User-Query $i$ and Document $d$ is described by the matrix $U(i,d)=U_{id}$. The value $U_{id}$ represents the satisfaction the User-Query $i$ obtains when presented with Document $d$. By construction the number of non-zero entries of the Utility matrix is:

$$NZU = (mn) / k \tag{2}$$

The simplest relation of the Utility, Document and User-Query matrices is the following [21]:

$$U = \sum_{t=1}^{k} R_t^T D_t \tag{3}$$

Each element $U_{id}$ of the Utility matrix is:

$$U_{id} = \sum_{t=1}^{k} R_{it} D_{td} \tag{4}$$

The elements $D_{td}$, $R_{ti}$ and $U_{id}$ define a tri-graph with subgraphs the Documents, the Topics and the Users-Queries (Diagram 1). In particular, the bipartite subgraph ([m], [n], L) of Document endorsements by Users-Queries is called by KKPS *the www state*.

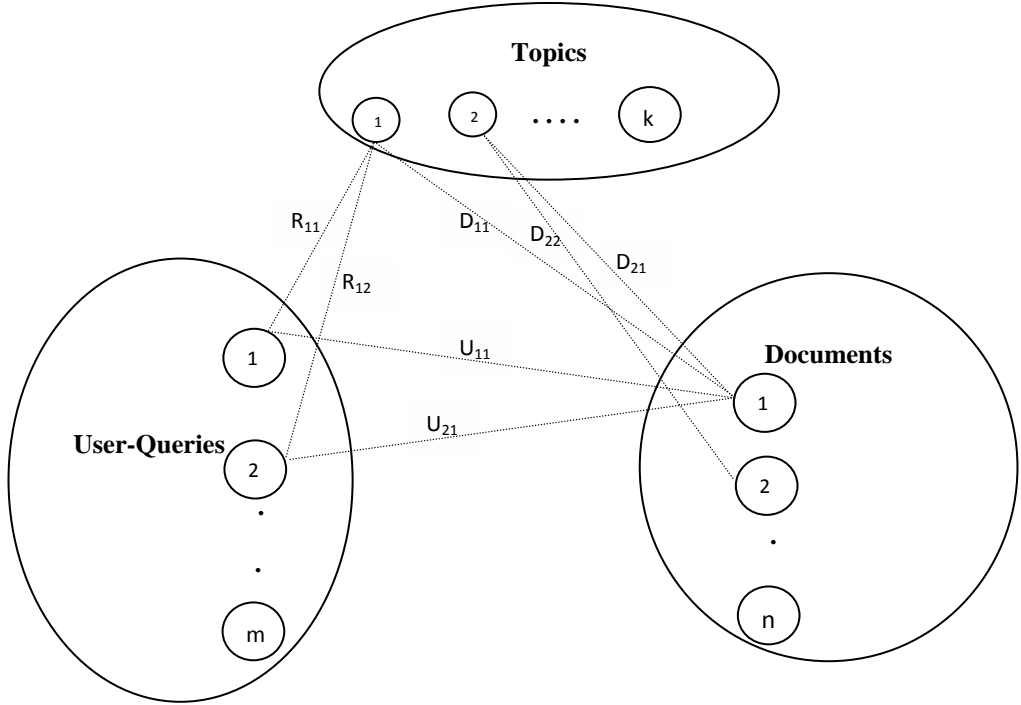

Figure 1: The Users-Queries, Documents, Topics tri-graph

The Search Engine proposes a number of Documents to the Users-Queries, Users-Queries choose and endorse those that have the highest Utility for them, and then Search Engine makes better recommendations based on these endorsements. It is assumed for simplicity that the number of Documents proposed by the Search Engine is fixed and denoted by $a$, and that the number of endorsements per User-Query is also fixed and denoted by $b$. It is assumed that:

$$b \leq a \leq n \tag{5}$$

Apparently, we have that:
- $m \cdot n$ is the maximum endorsements (complete bi-partite graphs),

- *m* is the maximum in-degree of a Document and the maximum average degree of the www state.

In the case of the KKPS model it is unlikely to reach maximum in-degree of a Document and the maximum average degree of the www state, because the Utility matrix restrictions are limiting the endorsement mechanism.

### 2.2. The algorithm

The structure of the *www state* emerges according to the following algorithm:

*Step 1*: A User-Query, for a specific Topic, is entered in the Search Engine.

*Step 2*: The Search Engine recommends *α* relevant Documents. The listing order is defined by a rule. In the very first operation of the Search Engine the Documents the rule is random listing according to some probability distribution.

*Step 3*: Among the *α* recommended Documents, *b* are endorsed on the basis of highest Utility. In this way, the bipartite graph *S*= ([*m*], [*n*], *L*) of Document endorsements is formed. Compute the in-degree of the Documents from the endorsements.

*Step 4*: Repeat Step 1 for another Topic.

*Step 5*: Repeat Step 2. The rule for Documents listing is the decreasing in-degree for the specific User-Query computed in Step 3.

*Step 6*: Repeat Step 3.

*Step 7*: Repeat Steps 4, 5, 6 for a number of iterations necessary for statistical convergence ("that is, until very few changes are observed in the www state" [21]).

### 2.3. Analysis

The basic assumptions of the evolution mechanism in the KKPS model are:
   a. every Document is relevant to at least one and even more than one Topic
   b. more than one Users-Queries could be relevant to the same Topic, resulting Documents with in-degree greater than one,
   c. a User-Query is concerned with strictly one Topic.

By construction the Utility Matrix has $m·n/k$ (2) non-zero entries, hence the final bipartite graph between Users-Queries and Documents has at most $m·n/k$ links (since a link between a User-Query and a Document is possible only if their Utility indicator is non-zero).

Additionally, since the out-degree for each User-Query equals $b$, the total out-degree of each iteration is $m \cdot b$. Therefore,

$$m \cdot b \leq m \cdot (n/k) \Longrightarrow b \leq n/k \quad or \quad k \leq n/b \tag{6}$$

Since in a bi-graph the total in-degree equals the total out-degree [22], we can assume that the maximum possible average degree is:

$$(m \cdot (n/k))/n = m/k \tag{7}$$

If each endorsement happens to be different from any other (each endorsement results a new link), then we have $r \cdot m \cdot b$ total endorsements, where r is the number of iterations. So,

$$r \cdot m \cdot b \leq m \cdot (n/k) \Longrightarrow r \leq n/(k \cdot b) \tag{8}$$

## 3. Results on the Power-law Exponent

Concerning the dependence of the power-law exponent on the number $α$ of recommended Documents by the Search Engine, the number $k$ of topics and the number $b$ of endorsed documents per User-Query, [21] found that for a wide range of values of the parameters $m$, $n$, $k$, $a$, $b$, the in-degree of Documents is power-law distributed. They emphasized that "more work is needed in order to define how the various parameters affect the exponent of the distribution." In order to clarify this dependence, we extend here the investigation in the following directions:

1. for Uniform, Poisson and Normal initial random distribution of Documents in-degree (Step 2) and
2. for different values of $α$, $b$ and $k$.

The results of the simulations are summarized in Figures 2 and 3. The corresponding simulation software program is provided in webscience.gr/wiki/index.php?title=Projects.

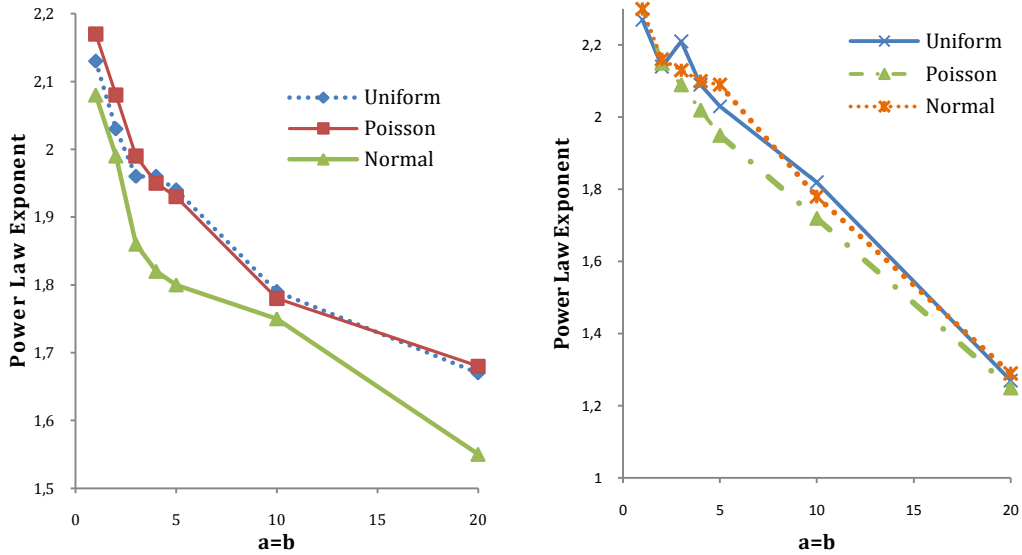

Figure 2: (i) in the case α=b, the validity of the power law becomes less significant as *b* increases for Uniform, Poisson and Normal initial random distribution of Documents in-degree (α=b=1, 3, 5, 10, 20 and k=80, m=750, n= 1500), (ii) an increase in the number of Topics *k*, results faster decay of the power law exponent (α=b=1, 3, 5, 10, 20 and k=120, m=750, n= 1500).

In the case of *α=b* (Figure 2), endorsements are formed only in the basis of the highest in-degree (Step 5), because there are no Documents left to be selected according to their highest Utility (Step 3). Setting the highest in-degree as the unique criterion for endorsements, means that Documents with high in-degree attract more endorsements. This is of course a preferential attachment [12] mechanism which weakens as *b* is increasing, because Documents with low in-degree are also endorsed (Figure 2 (i)). In addition, an increase in the number of Topics *k*, results in a faster decay of the power law exponent (Figure 2 (ii)).

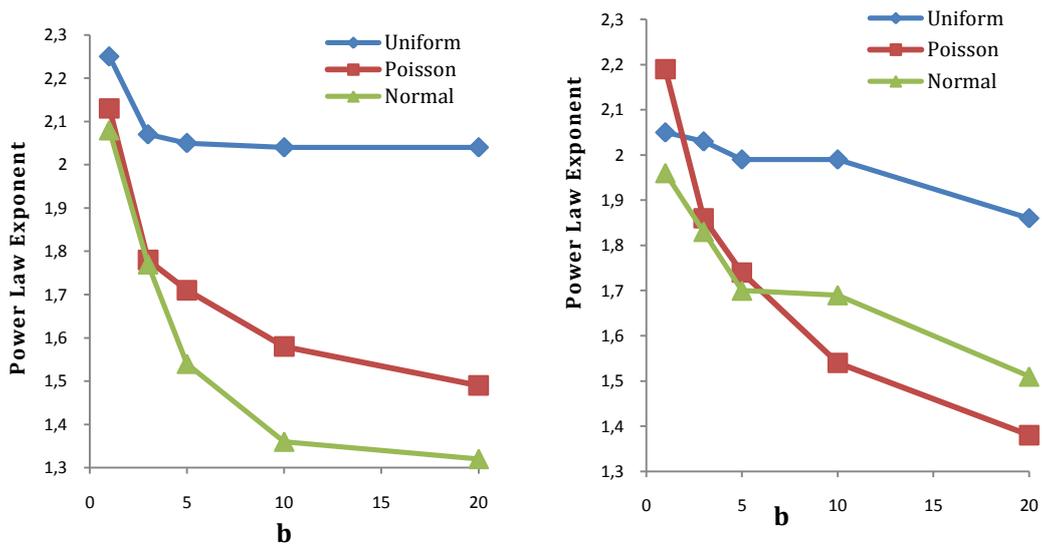

Figure 3: (i) in the case b ≤ $\alpha$, the validity of the power law becomes less significant as b increases for Uniform, Poisson and Normal initial random distribution of Documents in-degree (α= 20, b=1, 3, 5, 10, 20 and k=80, m=750, n=1500), (ii) An increase in the number of Topics k, results in faster decrease of the power law exponent (α= 20, b=1, 3, 5, 10, 20 and k=120, m=750, n=1500).

[7]

In the case $b \leq \alpha$, only the $b$ highest Utility Documents are selected from $a$ highest in-degree Documents. The validity of the power law becomes less significant as $b$ increases for Uniform, Poisson and Normal initial random distribution of Documents and an increase in the number of Topics $k$, results in a faster decrease of the power law exponent (Figure 3). For a wide range of values $m$, $n$, $k$, $a$, $b$ simulated results follow similar patterns.

## 4. Results on the "Price of Anarchy"

Efficiency or "price of anarchy" of the Search Algorithm is defined by [21] as the fraction of the maximum possible Utility (the ideal situation where each user sees the Documents of maximum Utility of him or her) that can be realized by a Search Engine. Concerning the dependence of the efficiency of the search algorithm on the number $\alpha$ of recommended Documents by the Search Engine, the number $k$ of topics and the number $b$ of endorsed documents per User-Query KKPS found that:

1. The efficiency of the algorithm increases, as the number $\alpha$ of recommended Documents by the Search Engine and the number $k$ of topics increase, however,
2. The efficiency of the algorithm increases, as the number $b$ of endorsed Documents per User-Query decreases.

Our examination of the issue confirmed the first result (Figures 4, 5, 6), but not the second one (Figure 7). Commenting on their result [21] remarked that "This is quite unexpected, since more user endorsements mean more complete information and more effective operation of the search engine. But the opposite happens: more endorsements per user seem to confuse the search engine." Our result (Figure 7) confirmed their intuition but not their simulation.

It is remarkable that in all our experiments (including Uniform, Poisson and Normal initial random distribution of Documents in-degree) the price of anarchy improved dramatically during the first 2-3 iterations while later the improvement was very slow.

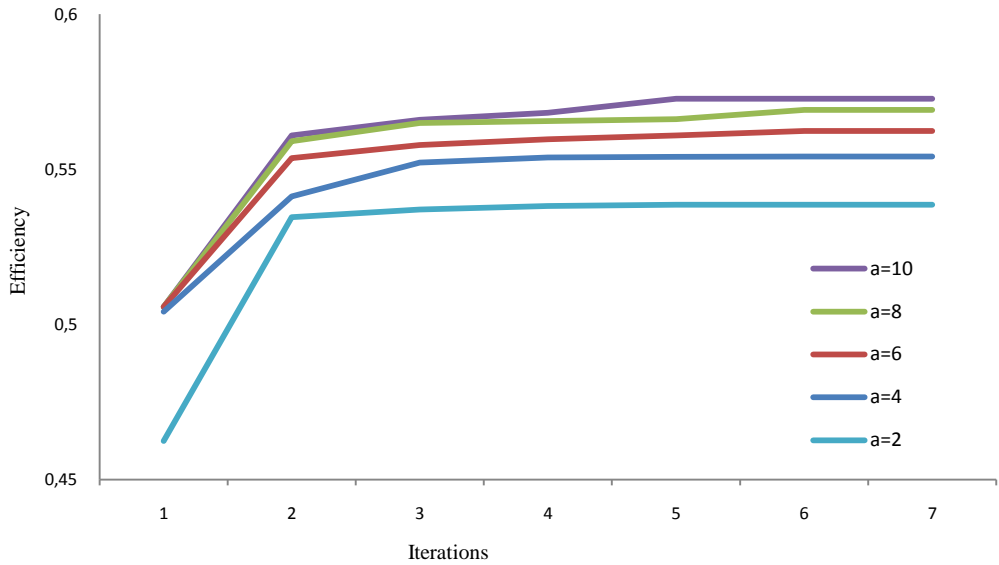

figure 4: (i) in the case b ≤ $\alpha$ efficiency of the search algorithm increases when the number of recommended Documents by the Search Engine $\alpha$ increases ($\alpha$=2, 4, 6, 8, 20, b=2, k=80, m=750, n=1500)

The efficiency of the algorithm increases, as the number $\alpha$ of recommended Documents by the Search Engine increases (Figure 4). This result was expected, since, for small values of $\alpha$, the algorithm proposes the highest in-degree but not necessarily the Documents with highest Utility because when $\alpha=b$ only the in-degree criterion is taken into account. As $\alpha$ increases there are more proposed Documents for the User-Query to endorse, thus the Utility criterion becomes more important than the in-degree criterion. The rate of improvement of the efficiency is decreasing as we approaching the absolute efficiency.

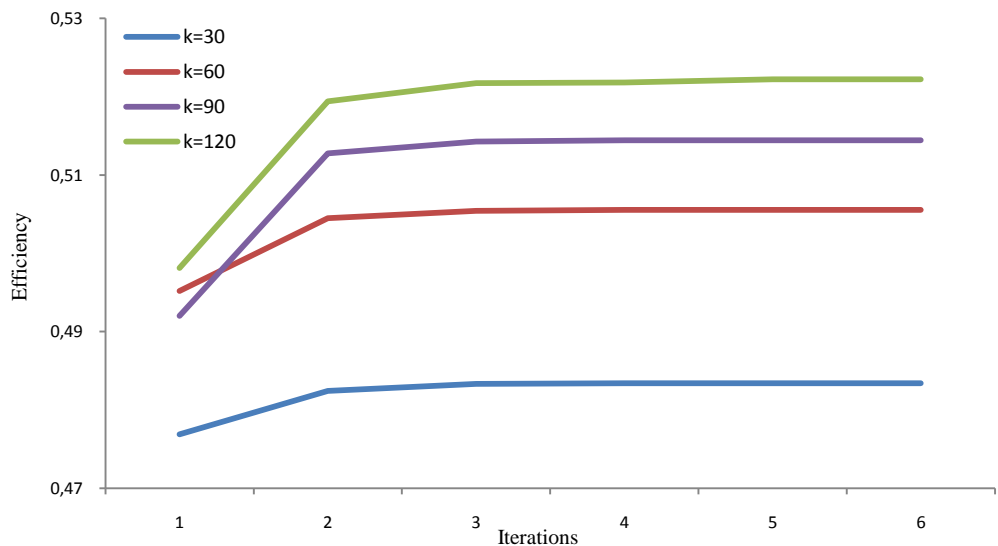

Figure 5: in the case $\alpha$=b efficiency of the search algorithm increases when the number of topics k increases ($\alpha$=b=1, k=30, 60, 90, 120, m=750, n=1500)

[9]

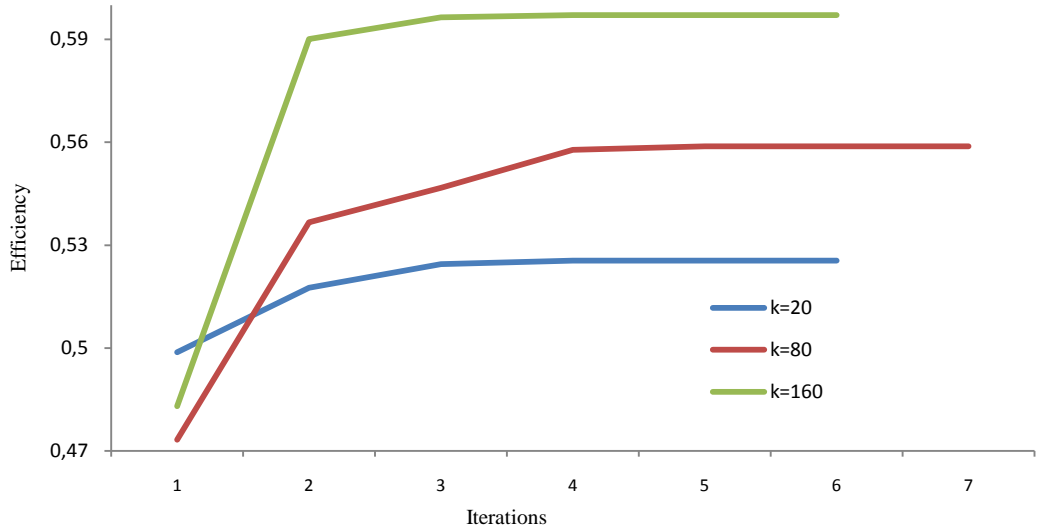

Figure 6: in the case b ≤ α efficiency of the search algorithm increases when the number of topics k increases (α=8, b=2, k=20, 80, 160, m=750, n=1500)

The efficiency of the algorithm increases, as the number *k* of Topics increases (Figures 5 and 6). This can be explained as follows. Efficiency is defined as the fraction U/TU, where U is the utility attained in each iteration and TU the total possible utility. By construction each vector D has *n/k* non zero elements and each vector R has *m/k* non zero elements. For smaller values of *k* (and constant values of the other parameters) the number of non zero elements of D and R vectors grows, causing the total attainable utility (TU) to grow instantly. On the contrary, the nominator of fraction U/TU grows slowly until becoming constant. Thus, the efficiency of the algorithm decreases for smaller values of *k*. Because of the absence of the Utility criterion in the case *α=b*, we observe the same results for different values of *k* (Figures 6).

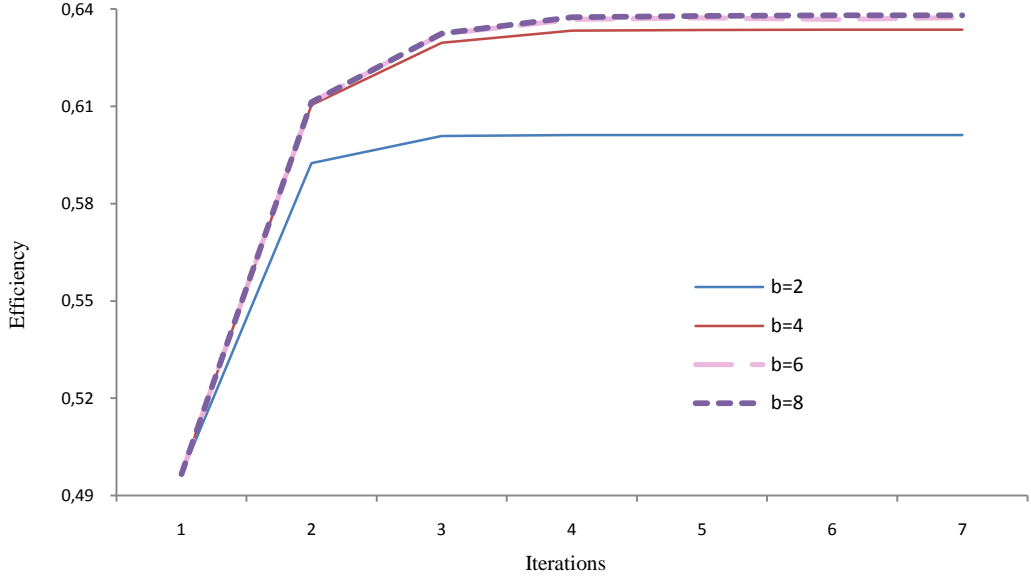

Figure 7: In the case b ≤ $\alpha$ efficiency of the search algorithm increases when the number of topics k increases (α=8, b=2, 4, 6, 8, k=160, m=750, n=1500)

The efficiency of the algorithm increases, as the number *b* of endorsed Documents per User-Query increases (Figure 7). This result could be explained by the following reasoning. When *b* increases, the utility U attained in each iteration, also increases. The algorithm (driven by the in-degree criterion) proposes *α* Documents for the User-Query and endorses *b* of them with the highest Utility. If the number of potential endorsements is bigger, let's say *b´*, the User-Query will certainly endorse the former *b* Documents, which have the highest Utility. Thus, the User-Query will enjoy maximal Utility (the Utility from endorsing *b* Documents, plus the Utility provided by the next *b´- b* Documents). This increases the efficiency of the algorithm for bigger values of *b*.

## 5. Discussion

1. Concerning the dependence of the power-law exponent on the number *α* of recommended Documents by the Search Engine, the number *k* of topics and the number *b* of endorsed documents per User-Query, we found that the validity of the power law becomes less significant as *b* increases, both in the case *α=b* and in the case *b ≤ a*, confirming the results of Kouroupas et al [21]. Our simulations however, extended the investigation for different initial random distributions of the in-degree of Documents and for different values of *α* and *b* (Section 4).

2. In the case *α=b*, Utility is useful only in terms of establishing compatibility between Utility Matrix and the Users-Queries and Documents bipartite graph, since all recommended Documents are endorsed according to the highest in-degree criterion.
3. Concerning the origin of the power law distribution of the in-degree of Documents, two mechanisms are identified in the KKPS model:
    - Users-Queries endorse a small fraction of Documents presented (*b*).
    - Assuming a small fraction of poly-topic Documents, the algorithm creates a high number of endorsements for them.

    The above mechanisms are not exhaustive for the real Web graph. Indexing algorithms, crawler's design, Documents structure and evolution should be examined as possible additional mechanisms for power law distribution.
4. Concerning the dependence of the efficiency of the search algorithm (price of anarchy [21]) on the number *α* of recommended Documents by the Search Engine, the number *k* of topics and the number *b* of endorsed documents per User-Query we found that the efficiency of the algorithm increases, as the number *α* of recommended Documents by the Search Engine, the number *k* of topics and the number *b* of endorsed Documents per User-Query increase (Section 5). Our simulations confirmed the results of Kouroupas et al [21], except the dependence on the number *b* of endorsed documents per User-Query where they found that "increasing *b* causes the efficiency of the algorithm to decrease. This is quite unexpected, since more user endorsements mean more complete information and more effective operation of the search engine. But the opposite happens: more endorsements per user seem to confuse the search engine." Therefore, in this case our result (Figure 7) confirmed their intuition but not their simulation.
5. According to [21] "The endorsement mechanism does not need to be specified, as soon as it is observable by the Search Engine. For example, endorsing a Document may entail clicking it, or pointing a hyperlink to it." This hypothesis does not take into account the fundamental difference between clicking a link (browsing) and creating a hyperlink. Clicking a link during browsing is the "temporal" process called traffic of the Web sites [23]. Web traffic is observable by the website owner or administrator through the corresponding log file [24] and by third parties authorized (like search engine cookies which can trace clicking behavior [25] or malicious. On the contrary, creating a hyperlink results in a more "permanent" link between two Documents which is observable by all Users-Queries and Search Engines. Therefore, the KKPS algorithm actually examines the Web traffic and not the hyperlink structure of Documents which is the basis of the in-degree Search engine's algorithm.

6. In this context, we remark that according to the published literature, Web traffic as well as Web content editing, are not taken into account in the algorithms of Search engines based on the in-degree (i.e. Pagerank [26]). These algorithms were built for Web 1.0 where Web content update and traffic monetization were not so significant. In the present Web 2.0 era with rapid change [27], the Web graph, content and traffic should be taken into account in efficient search algorithms. Therefore, birth-death processes for Documents and links and Web traffic should be introduced in Web models, combined with content update (Web 2.0) and semantic markup (Web 3.0 [28]) for Documents.
7. The discrimination between Users and Queries could facilitate extensions of the KKPS model in order to incorporate teleportation (a direct visit to a Document which avoids Search Engines) to a Document, different types of Users and relevance feedback between Documents and Queries [29].
8. In the KKPS model, Utility is defined to be time invariant linear function of $R$ and $D$ which by construction is not affecting the www state when $α=b$. This is a first approximation which does not take into account the dynamic interdependence of the Utility on the www state. In reality, the evolution of the www state will change both $R$ and $D$. A future extension of KKPS model should account for user behavior by incorporating Web browsing and editing preferences.
9. Lastly, it would be very useful to offer deeper insight in the Web's business model by incorporating economic aspects in the KKPS model. This could be achieved by introducing valuation mechanisms for Web traffic and link structures and monetizing the search procedure (sponsored search [30]).

## 6. Conclusions

The Web is the largest human information construct in history. Web technologies have been proven to be an enormous stimulus for market innovation, economic growth, social discourse and participation, and the free flow of ideas. The Web is transforming the society but how can we understand, measure and model its evolution in order to optimize its social benefit? Following the traditional wisdom of science we need realistic mathematical models which incorporate the salient features of the Web dynamics, namely structure, function and evolution. These include: the Web graph changes, the Web traffic, the Web content update, semantic processing, the distinction between Users and Queries and economic aspects.

The KKPS model of the fundamental Web components invests on a metaphor of the economic concept of utility. Our analysis clarified and extended the results of the KKPS model and highlighted future research developments in Web modeling which is the backbone of Web Science.

## 7. Acknowledgements

Fruitful discussions with Professor George Metakides are gratefully acknowledged. The Municipality of Veria, the authorities of the Aristotle University of Thessaloniki, the Faculty of Exact Sciences and the Mathematics Department supported enthusiastically the initiation of the Graduate Program in Web Science in Greece and our research.